\documentclass[conference]{IEEEtran}
\usepackage{amsmath, mathtools, amssymb, amsfonts, amsthm}
\usepackage{enumerate}
\usepackage{subfigure}
\usepackage{algorithm, algpseudocode}
\usepackage[thinc]{esdiff}
\usepackage{courier}
\usepackage{url}
\usepackage{adjustbox,multirow}
\usepackage{bbm, dsfont, csquotes}

\newcommand{\Adj}{\text{Adj}}

\newcommand{\D}{\mathsf D}
\newcommand{\R}{\mathsf R}

\newcommand{\Prob}{\mathbb P}
\newcommand{\RN}[1]{\textup{\uppercase\expandafter{\romannumeral#1}}}

\newtheorem{thm}{Theorem}[section]
\newtheorem{defn}{Definition}[section]
\newtheorem{lem}[thm]{Lemma}

\begin{document}

\title{Efficient Privacy-Preserved Processing of Multimodal Data for Vehicular Traffic Analysis}

% author names and affiliations, use a multiple column layout for up to three different affiliations
\author{\IEEEauthorblockN{Meisam Mohammady}
\IEEEauthorblockA{Iowa State University\\
meisam@iastate.edu}
\and
\IEEEauthorblockN{Reza Arablouei}
\IEEEauthorblockA{Data61, CSIRO\\
reza.arablouei@csiro.au}
}

\IEEEoverridecommandlockouts
\makeatletter\def\@IEEEpubidpullup{6.5\baselineskip}\makeatother
\IEEEpubid{\parbox{\columnwidth}{
    Symposium on Vehicles Security and Privacy (VehicleSec) 2023 \\
    27 February 2023, San Diego, CA, USA \\
    ISBN 1-891562-88-6 \\
    https://dx.doi.org/10.14722/vehiclesec.2023.23052 \\
    www.ndss-symposium.org
}
\hspace{\columnsep}\makebox[\columnwidth]{}}

\maketitle

\begin{abstract}

We estimate vehicular traffic states from multimodal data collected by single-loop detectors while preserving the privacy of the individual vehicles contributing to the data. To this end, we propose a novel hybrid differential privacy (DP) approach that utilizes minimal randomization to preserve privacy by taking advantage of the relevant traffic state dynamics and the concept of DP sensitivity. Through theoretical analysis and experiments with real-world data, we show that the proposed approach significantly outperforms the related baseline non-private and private approaches in terms of accuracy and privacy preservation.

\end{abstract}

\section{Introduction}

Differential privacy (DP) is commonly used in privacy-enhancing technologies, e.g., see~\cite{DBLP:conf/icdm/AcsCC12}-\cite{RastogiN10} and the references therein. However, utilizing DP-based techniques can pose certain challenges in a variety of applications. Examples are applying DP to set-valued datasets, which requires using a context-free taxonomy tree~\cite{chen11}, or to relational datasets, which is called non-interactive DP and usually done via sampling~\cite{leoni12}.

Datasets containing multimodal data appear in many real-world applications, e.g., regarding vehicular or network traffic or finance. Records in multimodal datasets are usually represented by value pairs such as $\langle x,y\rangle$ where $x$ is the modal value, e.g., the number of cars in an area, and $y$ is the mode, e.g., free or congested traffic. The straightforward application of DP to multimodal datasets requires randomizing both modal and mode values, which generally leads to significant sacrifice of accuracy.

In this paper, we present a novel hybrid DP approach that minimizes the required randomization through leveraging the underlying application-specific dynamics and the notion of sensitivity in DP, which is defined as the impact of changing the value of one data element over the outcome of a given query. We observe that multimodal dynamics often do not warrant randomizing the associated mode values as the DP sensitivity is not large enough to alter the mode. For instance, in vehicular traffic flows, adding or removing a single car can change the mode only at a transition point between the free to congested modes. Therefore, to apply DP in a traffic state estimation application, we can define two zones, namely, the safe zone and the sensitive zone, where the former contains the traffic states where DP randomization is not required for the modes, and the latter contains the states that require randomizing the modes.

To illustrate the efficacy of the proposed hybrid DP approach, we consider a traffic monitoring application. The corresponding dataset is collected by 27 single-loop detectors installed at various locations of the US Interstate 80 highway. This dataset comprises 18 hours of $\langle$count$,\ $occupancy$\rangle$ pair values recorded over a day. The field “count” is the number of cars passing by a detector and “occupancy” is the time during which the sensor is activated. Each pair value is recorded over a 30s continuous time window and the occupancy values are given as fractions of the associated 30s periods. The problem of interest is specifically the estimation of traffic state from the observed data while preserving the privacy of individual vehicles. The dynamics of traffic state can be represented by a nonlinear stochastic state-space model where the state is the density of the cars on the road and the observation data is the flow of the traffic.

We verify the effectiveness of the proposed hybrid DP approach both theoretically and empirically by comparing its performance to that of three benchmark approaches. The considered benchmarks are (i) a baseline non-privacy-preserving (non-private) approach that utilizes an extended Kalman filter where its prediction block optimally computes trustworthy state values from count and occupancy values, (ii) a baseline DP approach that applies two randomization mechanisms, i.e., an additive Gaussian randomization of counts and an exponential randomization of mode values \cite{7944526}, and (iii) a variant of the proposed approach that applies exponential DP mechanism only at a defined sensitive zone.

We show that the privacy guarantee of the considered baseline DP approach depreciates as the number of sensors grows. Intuitively, more records aggregated by the sensors mandate exponentially stronger exponential DP mechanism for mode values to maintain the privacy guarantee due to the increase in the sensitivity value. Our proposed hybrid DP approach for multimodal data addresses this dependency through carefully examining the state dynamics and determining the sensitivity according to the traffic status.

\section{Traffic Flow Dynamics} \label{sec:LWR TRAFFIC FLOW DYNAMICS}

The unidirectional traffic along a single road section, with the position denoted by $x$ and the varying number of lanes by $\lambda(x), $ can be mapped based on the traffic flow dynamics \cite{ketab} as $q=\rho v$ where $\rho$ is the vehicle density (e.g., in vehicles per mile) over all lanes, $q$ is the traffic flow over all lanes, and $v$ is the traffic velocity. Here, we assume the simple case of a homogeneous road section \cite{ketab} with the associated continuity equation expressed as
\begin{equation}
\label{eqn:partial}
\frac{\partial \rho  }{\partial t}+\frac{\partial (\rho v)  }{\partial x}=0.
\end{equation}
We consider a discrete version of \eqref{eqn:partial} by dividing the road section into cells of length $\Delta x_i$ and using a time step of $\Delta t$ \cite{ketab}. Hence, the density in cell $i$ over all lanes follows the recursion
\begin{align} \label{eqn:discretize}
\rho^i&(t+\Delta t)=\rho^i(t)\notag\\
&+\frac{\Delta t}{\Delta x^i}\left[F_{\text{tot}}\left(\rho^{i-1}(t),\rho^{i}(t)\right)-F_{\text{tot}}\left(\rho^{i}(t),\rho^{i+1}(t)\right)\right]
\end{align}
where $F_{\text{tot}}\left(\rho^{i-1}(t),\rho^{i}(t)\right)$ is the total numerical flux that enters cell $i$ (i.e., through the interface $i-1 \rightarrow i$) during period $\Delta t$, and $F_{\text{tot}}\left(\rho^{i}(t),\rho^{i+1}(t)\right)$ is the total numerical flux out of cell $i$ (i.e., through the interface $i \rightarrow i+1 $). Note that the numerical flux $F_{\text{tot}}\left(\rho^{i}(t),\rho^{i+1}(t)\right)$ is in general different from the total flow $q(x_{i|i+1},t)$, where $x_{i|i+1}$ denotes the location of the interface between cells $i$ and $i+1$. More details are provided in the following.

To complete the model, we need to make a hypothesis on the relationship between two quantities, e.g., velocity and density, or flow and density. Thus, we first introduce lane-averaged (also called effective) quantities, i.e., lane-averaged traffic density $\rho(x,t)$ (say, in vehicles per mile per lane), lane-averaged traffic speed $v(x,t)$, and lane-averaged traffic flow $q(x,t)=\rho(x,t)v(x,t)$ \cite{ketab}. Denoting by $\rho_j(x,t)$, $q_j(x,t)$, and $v_j(x,t)$ the density, speed, and flow in lane $j$ at position $x$, we have the relations
\begin{align*}
\rho(x,t)&=\frac{\sum^{\lambda(x)}_{j=1}\rho_j(x,t)}{\lambda(x)}\\
q(x,t)&=\frac{\sum^{\lambda(x)}_{j=1} q_j(x,t)}{\lambda(x)}\\
v(x,t)&=\frac{\sum^{\lambda(x)}_{j=1}v_j(x,t)}{\lambda(x)}.
\end{align*}

We adopt a first-order model or fundamental diagram, considering a static relationship $q(\rho)$. In first-order models, proposed in \cite{lwr1,lwr2} and known as Lighthill-Whitham-Richards (LWR) models, the effective density is a fundamental quantity and a sufficient description of the local traffic state, since the effective speed and the effective flow are assumed to be known static functions of density. The LWR models assume that the traffic flow is always in local equilibrium with respect to the density. This may lead to the formation of physically impossible phenomena such as shock waves. Regardless, the LWR models are widely used for modeling traffic flow dynamics~\cite{ketab}. Here, we use a triangular fundamental diagram as our LWR model. In the following, we describe this model, which is also called the cell-transmission model (CTM) \cite{asaai}.

\subsection{Cell Transmission Model}

\begin{figure}
\centering
\includegraphics[width=0.6\linewidth]{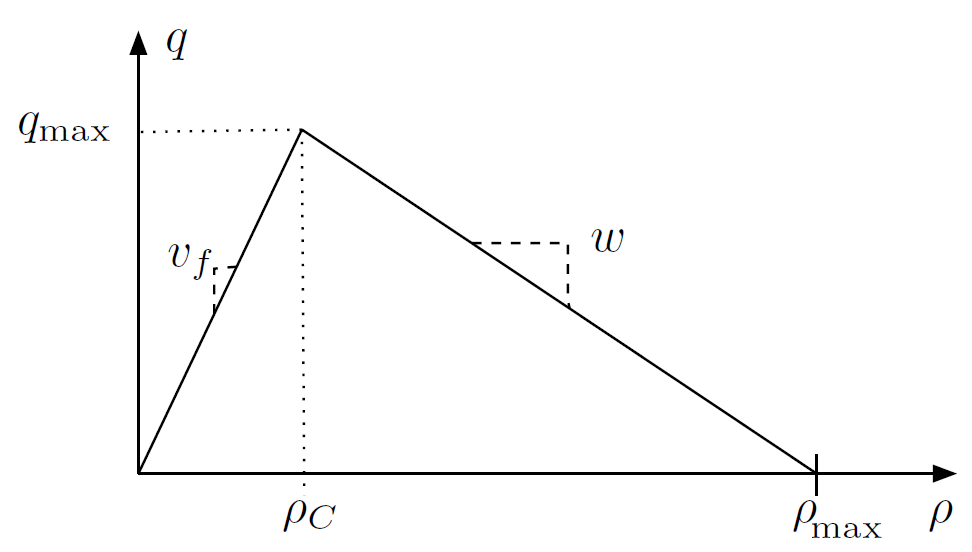}
\caption{Triangular fundamental diagram and the associated parameters.}
\label{fig:LWR1}
\end{figure}

The CTM is a simple LWR model that uses a triangular fundamental diagram formulated as
\begin{equation}
q(\rho)=\begin{cases}
v_f \rho \qquad \qquad \quad if \quad \rho\leq \rho_c \\ 
w(\rho_{\max}-\rho) \quad if \quad \rho_c \leq \rho \leq\rho_{\max}. \\
\end{cases}
\label{eqn:lwr}
\end{equation}
Here, $v_f$ is the velocity of free traffic (say 110 km/h for a highway), $\rho_{\max}$ is the maximum density on this road segment (say $120$ vehicles/lane/km for a highway), $\rho_c$ is the critical density at which the maximum flow $q_{\max} = v_f \rho_c$ is attained, and $w$ is the velocity of the waves of density variations in congested traffic that propagate backwards. Fig. \ref{fig:LWR1} illustrates these definitions \cite{ketab}.

Dividing the road into $I$ cells numbered as $1,...,I$, the discrete-time lane-averaged conservation law for vehicles corresponding to the solution $q(\rho)$ of \eqref{eqn:partial} is
\begin{align*}
\rho^i_{k+1}=\rho^i_{k}+\frac{\Delta t}{\Delta x_i}\left(\frac{\lambda^{i-1}}{\lambda^i}F(\rho^{i-1}_{k},\rho^{i}_{k})-F(\rho^{i}_{k},\rho^{i+1}_{k})\right),
\end{align*}
for $i=1,...,I$, where $\rho^{i}_{k}$ is the lane-averaged vehicle density in cell $i$ at period $k$, i.e., during the time interval $\left[ k \Delta t, (k + 1)\Delta t \right]$, and $F\left(\rho^{i}_{k},\rho^{i+1}_{k}\right)$ is the lane-averaged numerical flux out of cell $i$, i.e., through the interface $i \rightarrow i+1 $, during period $k$. We also define $\lambda^i$ to be the number of lanes at the interface $i \rightarrow i+1$. Any location where the number of lanes changes is presumed to fall inside a cell. This leads to a system with non-linear (piecewise linear) dynamics. At the ends of the road for which we estimate the traffic, we add two ghost cells numbered $0$ and $I + 1$ to enforce the boundary conditions. In order to enforce the boundary conditions, we assume that there are loop detectors at the exit of cell $0$ and at the entrance of cell $I + 1$ \cite{asaai}. Therefore, we obtain the following stochastic state-space model of the density dynamics on the road:
\begin{equation}
\label{eqn:discrete}
\rho^i_{k+1}=\rho^i_{k}+\frac{\Delta t}{\Delta x_i}\left(\frac{\lambda^{i-1}}{\lambda^i}F(\rho^{i-1}_{k},\rho^{i}_{k})-F(\rho^{i}_{k},\rho^{i+1}_{k})\right)+ \gamma^i_k,
\end{equation}
for $i=1,...,I$. Here, $\gamma^i_k$ is a Gaussian random variable whose variance can be tuned in the design of the state estimator, based on the relative confidence we place in the model or the observations. The dynamics of the ghost cells are also expressed as
 \begin{equation}
\label{eqn:discrete2}
\rho^0_{k+1}=\rho^0_{k}+ \gamma^0_k,\ \ \rho^{I+1}_{k+1}=\rho^{I+1}_{k}+ \gamma^{I+1}_k.
\end{equation}
Finally, for the triangular fundamental \eqref{eqn:lwr}, the standard numerical method of Godunov corresponds to using the following numerical flux in \eqref{eqn:discrete} as
\begin{equation}
F(\rho^{i}_k,\rho^{i+1}_k)=\min\left(\rho^i_k v_f,\rho_c v_f,w(\rho_{\max}-\rho^{i+1}_k)\right).
\end{equation}

The stochastic state-space model \eqref{eqn:discrete} together with the measurements reported by single-loop detectors (see section \ref{sec:singleloop} ahead) can be assimilated into an extended Kalman filter (EKF) to construct a traffic density map. We briefly describe the EKF In Appendix II.

\section{Single-Loop Detector Measurement Model} \label{sec:singleloop}

The datasets provided by the flow sensors consist of sequences of counts $c^i_{j,k}$ and occupancies $o^i_{j,k}$ for $k\geq 0,1\leq i\leq S$, and $0\leq j\leq\lambda^i$. Here, $k$ indicates the related 30s period, $S$ is the number of single-loop detectors reporting the records, and $j$ is the lane number. The occupancy $0\leq o^i_{j,k}\leq 1$ is a unitless number representing the fraction of the $k$th period during which any vehicle has passed before sensor $i$. The single-loop detectors cannot directly measure the traffic density or velocity at their locations. However, their measurements can be used to estimate these quantities. For a single-lane road equipped with single-loop detectors, the estimates are
\begin{align}
\label{eqn:approx relation}
v_j(t) \approx g \frac{c_j(t)}{o_j(t)T}, \ q_j(t) \approx \frac{c_j(t)}{T}, \ \rho_j(t) \approx \frac{o_j(t)}{g}
\end{align}
where $T$ is the time period of the sensor (30 seconds here) and $g$ is the g-factor, which denotes the average effective vehicle length at the sensor location and can vary over time. Similar to \cite{asli}, to obtain a more robust approximation of density, we first compute the approximate flows based on the count data. That is, we express the flow $\phi^i_k$ around the sensor placed at the interface $i\rightarrow i+1$ for cells $i$ and $i+1$ by the following non-linear measurement model
 \begin{equation}
 \label{eqn:flows}
 \phi^i_k=\frac{1}{\lambda^i T} \sum^{\lambda^i }_{j=1} c^i_{j,k}=F(\rho^{i}_{k},\rho^{i+1}_{k})+\nu_k
 \end{equation}
where $\nu_k$ is a Gaussian random variable representing the measurement error or noise. We then define the density pseudo-measurement model as
\begin{equation}
\label{eqn:flowdensity}
z^i_k=z^{i+1}_{k}=\begin{cases}
\frac{\phi^i_k}{v_f} \qquad \qquad \ \text{if} \quad m^i_k=F\\ 
\rho_{\max}-\frac{\phi^i_k}{w} \quad \text{if}  \quad m^i_k=C\\
\end{cases}
\end{equation}
where $m^i_k$ denotes the traffic mode of the interface that is either free (F) or congested (C) corresponding to $\rho \leq \rho_c$ or $\rho > \rho_c$, respectively. This model can be obtained by inverting our triangular fundamental diagram presented in \eqref{eqn:lwr}. The observation signal $z^i_k$ is related to the density of the flow as
\begin{equation}
z^i_k=z^{i+1}_k=\rho^i_k+\eta^i_k=\rho^{i+1}_k+\eta^{i+1}_k
\end{equation}
where $\eta^i_k, \eta^{i+1}_k$ are assumed to be Gaussian random variables. This model requires determining the exact mode of the traffic flow. The strategy proposed in \cite{asli} is to use the reported occupancy measurements and estimate the traffic mode to be either fluid or congested based on whether $\frac{o_j}{g} \leq \rho_c$ or $\frac{o_j}{g} > \rho_c$, respectively.

The above measurements can result in frequent traffic mode estimation errors due to inaccurate approximation of g-factor. Given cars being at least 18 feet long and trucks being up to 60 feet long, the g-factor parameter is expected to range from 18 feet for inner car-only lanes to 60 feet in the early morning for outer lanes over fluid highways with heavy truck traffic. These mode measurements are difficult to handle from a DP point of view, because the occupancy time due to a single vehicle, denoted by $\frac{l_v}{T v_v}$ with $l_v$ being the length and $v_v$ the speed, can vary widely depending on its speed. As a result, the sensitivity of these occupancy measurements is high and the standard Gaussian perturbation mechanism exacerbates the reliability of the measurements, especially at low density \cite{asli}. We now present our mode measurement model, which takes both the occupancy and the count measurements into account to obtain a more reliable estimation of the traffic mode.

\section{Non-Private Mode and Density Measurements} \label{sec:nonpriv}

According to \eqref{eqn:flowdensity}, two possible densities on the fundamental diagram correspond to each flow measurement $0 \leq \phi_{k}^i < q_{\max}$ \eqref{eqn:flows}. Based on \eqref{eqn:approx relation}, we can also form the lane-average contribution to the density via occupancy measurements as 
\begin{align}
\label{eqn:occdensity}
y^{i}_k=\frac{1}{g \lambda^i} \sum^{\lambda^i}_{j=1}o^i_{j,k}.
\end{align}
The traffic mode pseudo-measurements can then be obtained as
\begin{align}
\label{eqn:closurenn}
M^{i}_k=M^{i+1}_k=\operatorname*{arg\,min}_{m^i_k} \left|z^i_k(m^i_k) -y^i_k \right|.
\end{align}

The model \eqref{eqn:closurenn} estimates the mode as either free (F) or congested (C) based on which subfunction in the hybrid function $z^i_k(m^i_k)$ \eqref{eqn:flowdensity} is closer to the occupancy contribution to the density $y^i_k$. This model requires an accurate estimate of the g-factor parameter to guarantee that the minimum in \eqref{eqn:closurenn} is correctly evaluated. This is challenging as g-factor can change over time and is generally not easy to estimate. To tackle this, we assume that g-factor is constant, namely $20$ feet. We then bound the allowed deviation between the density pseudo-measurement $z^i_k$ \eqref{eqn:lwr} and the occupancy contribution to density $y^i_k$ \eqref{eqn:occdensity}. That is, for constant $g$, $\exists \zeta(g)>0$ such that $\left|\log z^i_k- \log y^i_k\right| \leq \zeta(g)$ $\forall i,k$. This limits the variations in g-factor as
\begin{align}
\label{eqn:gfactorbaze}
%\left|\log \frac{z^i_k}{y^i_k}\right| \leq \zeta(g)  \equiv \ 
\frac{1}{g e ^{\zeta(g)} \lambda^i} \sum^{\lambda^i}_{j=1}o^i_{j,k} \leq z^i_k \leq \frac{e^{\zeta(g)}}{g\lambda^i} \sum^{\lambda^i}_{j=1}o^i_{j,k}.
\end{align} 

Given the assumed values of $g$ and $\zeta(g)$, we define the sets
\begin{align}
\label{T_F}
T_F= \left \{\left(\phi^i_k,y^i_k \right):\left|\log\tfrac{\phi^i_{k}}{v_f}-\log y^i_k \right| \leq \zeta(g)\ \forall i,k \right\}   
\end{align}
\begin{align}
\label{T_C}
T_C= \left \{\left(\phi^i_k,y^i_k \right):\left|\log\left(\rho_{\max}-\tfrac{\phi^i_{k}}{w}\right)-\log y^i_k\right| \leq \zeta(g)\ \forall i,k \right\}
\end{align}
corresponding to the flow $ \phi^i_k$ satisfying our truncation in the free and congested modes, respectively. Consequently, we introduce the following lemma.\footnote{We provide the proofs of all theorems and lemmas in Appendix III.}
\begin{lem} \label{thm:nonprivatezone}
For any flow $\phi^i_k$, defined in \eqref{eqn:flows}, we have  
\begin{align}
\label{eqn:safezone}
\mathbf{1}_{T_F}&\left((\phi^i_k,y^i_k)\right) \mathbf{1}_{T_C}\left((\phi^i_k,y^i_k)\right)=1\\
&\text{iff}\ \phi^i_k \in \left[\dfrac{w v_f \rho_{\max}}{w e^{2 \zeta(g)} +v_f},\dfrac{w e^{2 \zeta(g)} v_f \rho_{\max}}{w+e^{2 \zeta(g)} v_f} \right].
\end{align}
\end{lem}

\begin{center}
\begin{figure}
\centering
\includegraphics[width=1\linewidth]{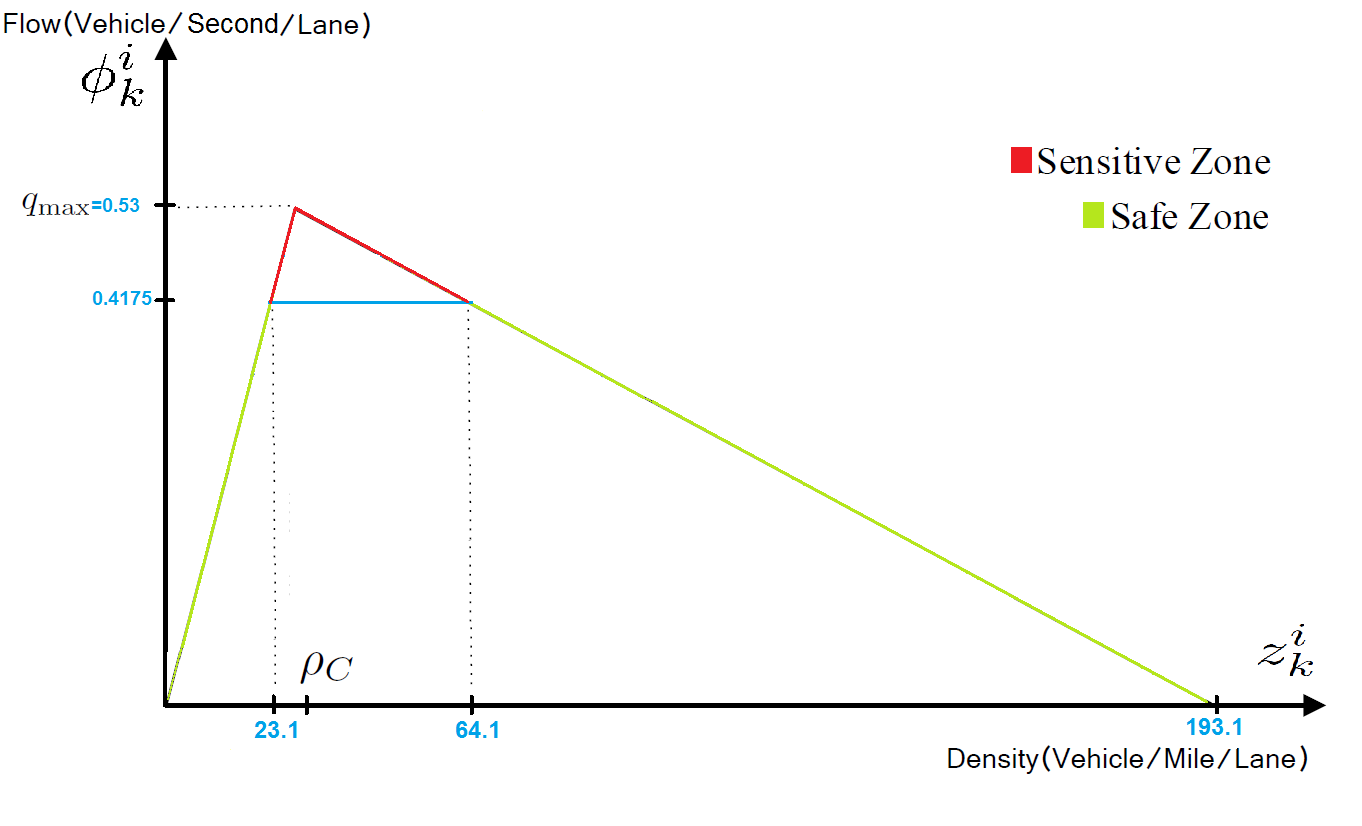}
\caption{Safe and Sensitive zones on a triangular fundamental diagram for $g=20$ feet and $\zeta(g) =0.51$.}% The safe zone is quite large for simulation purposes.}
\label{fig:LWR}
\end{figure}
\end{center}

Defining $\bar{T}_C$ and $\bar{T}_F$, the complement sets of $ T_C$ and $T_F$, respectively, we obtain a traffic mode measurement model as
\begin{equation}
\label{eqn:model1}
M^i_k=\begin{cases}
F \qquad \text{if} \ \mathbf{1}_{T_F-T_C}\left((\phi^i_k,y^i_k)\right)=1:\text{Safe zone, F mode}\\ 
C \qquad \text{if} \ \mathbf{1}_{T_C-T_F}\left((\phi^i_k,y^i_k)\right)=1:\text{Safe zone, C mode}\\
M^i_{k-r} \ \text{if} \ \left[\prod^{r-1}_{s=0} \mathbf{1}_{T_C \cap T_F}\left((\phi^i_{k-s},y^i_{k-s})\right) \right]\times\\
\mathbf{1}_{\bar{T}_C \cup \bar{T}_F}\left((\phi^i_{k-r},y^i_{k-r} )\right)=1, \ r>0:\text{Sensitive zone} 
\end{cases} 
\end{equation}
The mode measurement model \eqref{eqn:model1} determines the mode of the traffic as either free (F) or congested (C), if the current flow satisfies \eqref{eqn:model1} only in free mode or only in congested mode, respectively. The third case corresponds to the flow
\begin{equation*}
\phi^i_k \in \left[\dfrac{w v_f \rho_{\max}}{w e^{2 \zeta(g)} +v_f},\dfrac{w e^{2 \zeta(g)} v_f \rho_{\max}}{w+e^{2 \zeta(g)} v_f} \right]  
\end{equation*}
where the truncation is respected in both traffic modes. For this case, we take the mode of the last flow $\phi^i_{k-r}$, which is inside one of the two safe zones in \eqref{eqn:model1}. To illustrate this, we depict the region corresponding to the flows satisfying the truncation in one mode with green lines on a triangular fundamental diagram in Fig. \ref{fig:LWR}. We call the region corresponding to these flows, the \enquote{Safe} zone. We also introduce a \enquote{Sensitive} zone (red line), which represents the flows that our model is unable to determine their mode directly. The parameters related to the fundamental diagram and the model that are used in Fig. \ref{fig:LWR} are $v_f=65$ mph, $w=11.6$ mph, $\rho_{\max}=193$ vehicles/mile/lane, $g=20$ feet, and $\zeta(g) =0.51$. As per \eqref{eqn:gfactorbaze}, the choice of $\zeta(g) =0.51$ corresponds to g-factor variations between 12 and 33.3 feet.

Our mode measurement model \eqref{eqn:model1} estimates the mode of the traffic with respect to the flows falling inside the Safe zone. Although this strategy can reduce the accuracy of the mode measurement, especially for flows in the Sensitive zone, in section \ref{sec:privmodel}, we show that this model can efficiently be used to design a differentially-private traffic estimator.

To obtain a more physically meaningful mode, we filter the mode measurements through an additional hidden-Markov model (HMM), as explained below. For the state trajectory $M^i_k$ defined in \eqref{eqn:model1}, we utilize the actual mode estimate used to invert the fundamental diagram that is the new state trajectory $\{s^i_k\}_{k \geq 0}$ with $s^i_k \in \{C,F\}$. We describe the dynamics of $s^i_k$ via a Markov chain with a single parameter $\pi_1= \mathbb{P}(s^i_{k+1} \neq s^i_k)$, which represents the probability of mode changing from free to congested at that location. This parameter can be estimated from historical data. We introduce another parameter $\pi_2 = \mathbb{P}(m^i_k=s^i_k)$ that reflects our confidence in the output of our model. We set the confidence probability parameter in HMM with respect to the flow data as its values for the flows in the Sensitive zone ought to be lower compared to those for the flows in the Safe zone. For non-private estimation, we can define the confidence probability as $\pi_2 = \mathbb{P}(m^i_k=s^i_k| q^i_k)$, which is useful for Sensitive zone flows. Even if $ M^i_{k-r}$ addresses a wrong mode, the confidence probability is set according to the occupancy contribution to density \eqref{eqn:occdensity} hence HMM can correct the error. We summarize the procedure of providing density measurements from occupancies and counts data in Algorithm \ref{alg:1}.

\begin{algorithm}
\caption{Non-private density measurement.} \label{alg:1}
\begin{algorithmic}[1]
\State Calculate flow measurements $\phi^i_k=\frac{1}{T \lambda^i} \sum_{j=1}^{\lambda^i}
 c^i_{j,k}$.
\State Based on historical data, choose a base g-factor, e.g, 20 feet, and an upper-bound error $\zeta(g)$.
\State Specify the corresponding Safe and Sensitive zones based on Theorem \ref{thm:nonprivatezone}.
\State Calculate $m^i_k=\frac{F}{C}$ based on the mode measurement model \eqref{eqn:model1}.
\State Filter $m^i_k$ thorough the HMM filter to obtain the actual mode $s^i_k$ used to invert the fundamental diagram.
\State Calculate
 \begin{equation*}
z^i_k=z^{i+1}_{k}=\begin{cases}
\frac{\phi^i_k}{v_f}\qquad\qquad\ \text{if}\quad s^i_k=F\\
\rho_{\max}-\frac{\phi^i_k}{w}\quad \text{if}\quad s^i_k=C.\\
\end{cases}
\end{equation*}
\end{algorithmic}
\end{algorithm}

\begin{figure}
\centering
\includegraphics[width=0.7\linewidth]{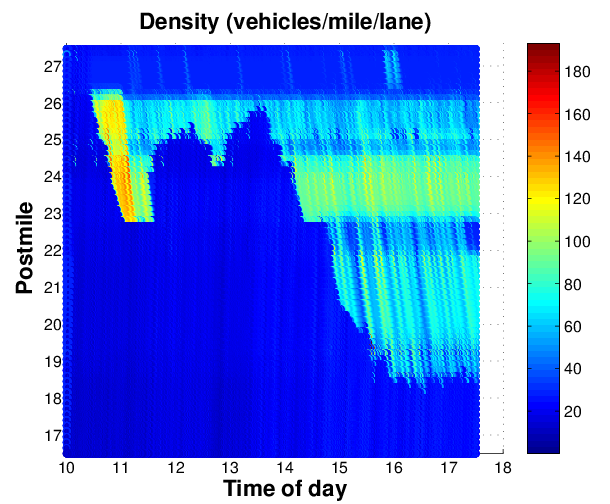}
\caption{Real-time density map reconstruction using a non-private EKF based on Algorithm \ref{alg:1}.}
\label{fig:nonprivatehmmfilterasli-eps-converted-to2}
\end{figure}

\begin{figure}
\centering
\includegraphics[width=0.7\linewidth]{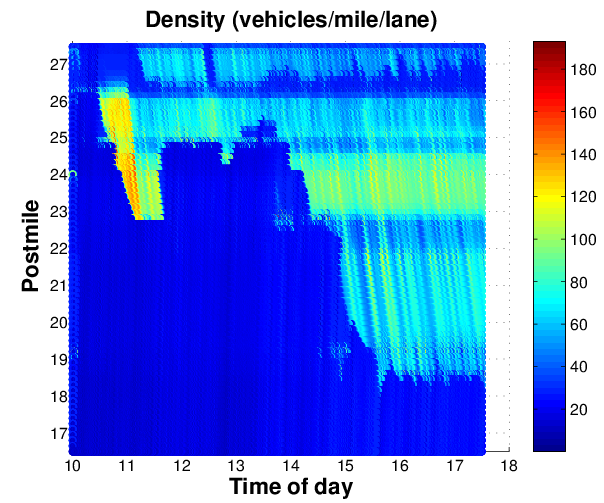}
\caption{Real-time density map reconstruction using a non-private EKF presented in \cite{asli}.}
\label{fig:nonprivatejerom}
\end{figure}

To illustrate our approach, we estimate the traffic state from induction loop data available as part of the Mobile Century dataset \cite{Herrera2010568}. This data consists of counts and occupancy measurements from single-loop detectors for each northbound lane of US Interstate $880$ highway between post-miles $16.5$ and $27.7$ (along an approximately $11$-mile-long road section). We assimilate the density measurements based on Algorithm \ref{alg:1} in an EKF to construct the non-private density map shown in Fig. \ref{fig:nonprivatehmmfilterasli-eps-converted-to2}. The resulting map is similar to the map in Fig. \ref{fig:nonprivatejerom} that is the non-private map produced by \cite{asli}. This similarity proves the reliability of our mode measurement model. The two maps have some discrepancies mainly in the areas where the traffic is about to switch between the modes. Fig. \ref{fig:nonprivatehmmfilterasli-eps-converted-to2} is likely a more reliable picture of the traffic density as our mode measurement model considers the possible variations of g-factor over time. In the following, we show that this model can be used efficiently in a differentially-private scheme.


\section{Differentially-Private Mode and Density Measurements} \label{sec:privmodel}

The measurements obtained from the single-loop detectors, i.e, counts $c^i_{j,k}$ and occupancies $o^i_{j,k}$, cannot be directly used in any traffic estimator architecture, because they may reveal private information about individuals who contribute to these measurements. In this section, we present deferentially private algorithms that output privacy-preserved flow and traffic mode measurements. These sanitized pseudo-measurements are then used to provide differentially-private density pseudo-measurements, which are sufficient to construct our differentially-private density map. We use a Gaussian mechanism, which was first presented in \cite{asli}, to process the counts data and provide differentially-private traffic flow data. We provide some background information on the notion of DP in Appendix I.

In order to preserve the privacy of traffic mode measurements, we introduce a new mechanism for sanitizing data sequences that is mainly based on Algorithm \ref{alg:1}. In what follows, we first review the Gaussian mechanism (Theorem \ref{thm: Gaussian mech}) providing the privacy-preserved flow measurements. Then, we present the mechanism that provides the mode pseudo-measurements.

\subsection{Flow Measurements}

We can consider the following adjacency relation for the counts datasets of $N$ user trajectories $ C:=\{c^i_{j,k}: k\geq 0, 1\leq i\leq S,1 \leq j \leq\lambda^i\}$ 
\begin{align} \label{eqn:adjac}
&\forall c,\tilde{c}\in C:\Adj(c,\tilde{c})\ \textrm{iff}\ \forall k \geq 0,\ \forall i \in \left[1,S \right], \forall j \in \left[1,\lambda^i \right],\notag\\
&i,j \in \mathbb{N},\ \exists \left(j_1,k_1 \right) , \left(j_2,k_2 \right):\left|c^i_{j_1,k_1}-\tilde{c}^i_{j_1,k_1} \right| \leq 1,\notag\\
&\left|c^i_{j_2,k_2}-\tilde{c}^i_{j_2,k_2} \right| \leq 1,\ c^i_{j,k}=\tilde{c}^i_{j,k}\ \forall \left(j,k \right) \neq \left(j_1,k_1 \right), \left(j_2,k_2 \right).
\end{align} 
This adjacency relation indicates that changing the trajectory of a single car can affect the counts measurements reported by each sensor $i$ in at most two different time steps. To make this clearer, suppose that Jane's car triggers a number of sensors every day when she goes to her job in the morning. For any DP mechanism to hide her absence or presence, we must keep in mind that changing her trajectory can change the reported counts of each sensor at two different times, one corresponding to a unit decrease in her usual trend of passing and the other corresponding to a unit increase in her new trend.

Let us consider two adjacent flow datasets as $\phi^i_k$, expressed in \eqref{eqn:flows}, and $ \tilde{\phi^i_k}$. Then, we have
\begin{align*}
\left\|\phi- \tilde{\phi} \right\|^2_2=\sum^{M}_{i=1} \sum^{\infty}_{k=0}  \left|\phi^i_{k}-\tilde{\phi}^i_k \right|^2.
\end{align*} 
For a sensor at the interface $i \rightarrow i+1$, the corresponding term is
\begin{align*}
\sum^{\infty}_{k=0} \left|\phi^i_{k}-\tilde{\phi}^i_k \right|^2=  \frac{1}{T^2 (\lambda^i)^2}\sum^{\infty}_{k=0}  \left|\sum^{\lambda_i}_{j=0} \left(c^i_{j,k}- \tilde{c}^i_{j,k} \right) \right|^2.         
\end{align*}  
Based on the adjacency relation \eqref{eqn:adjac}, the counts $c^i_{j,k}$ and $\tilde{c}^i_{j,k}$ must be almost all identical, except that some vehicles $A$ and $B$ can cross the line of the sensor at different periods and in different lanes. 
Thus, we have
\begin{align*}
\sum^{\infty}_{k=0} \left|\phi^i_{k}-\tilde{\phi}^i_k \right|^2 \leq \frac{2}{T^2 \left(\lambda^i \right)^2}
\end{align*}
and hence
\begin{align}
\label{eqn:countsens}
\left\|\phi- \tilde{\phi} \right\|^2_2= \sum^{M}_{i=1} \sum^{\infty}_{k=0}  \left|\phi^i_{k}-\tilde{\phi}^i_k \right|^2 \leq \frac{2}{T^2} \sum^M_{i=1} \frac{1}{\left(\lambda^i \right)^2}:\Delta f^2.         
\end{align} 
Now, given Theorem \ref{thm: Gaussian mech}, the mechanism that publishes the perturbed flow pseudo-measurements $\Phi^i_k=\phi^i_k+n^i_k$, where $ n^i_k$ are independent zero-mean white Gaussian noise with covariance $\kappa^2_{\delta,\epsilon}\Delta f^2$ and $\Delta f$ as in \eqref{eqn:countsens}, is ($\epsilon$, $\delta$)-differentially private.

\subsection{Density and Mode Measurements}

The flow pseudo-measurements obtained from the Gaussian mechanism, $\Phi^i_k=\phi^i_k+n^i_k$, can be used to calculate the density pseudo-measurements, but this requires an additional mode estimate. However, estimating the traffic mode based on the count/occupancy datasets and without sanitization can compromise private information of individuals. In this section, we present our privacy-preserved mode measurement that is mainly based on the mode measurement model presented in Section \ref{sec:nonpriv}. Like \eqref{eqn:adjac}, the adjacency relation for the occupancy data of $N$ user trajectories $O:= \{o^i_{j,k}:k\geq 0, 1 \leq i \leq S, 0 \leq j \leq \lambda^i\}$ is
\begin{align} \label{eqn:occadjac}
&\forall o, \tilde{o} \in O:\ \Adj \left(o,\tilde{o} \right)\ \textrm{iff}\ \forall k \geq 0, \forall i \in \left[1,S \right], \forall j \in \left[1,\lambda^i \right],\notag\\
&i,j \in \mathbb{N}, \exists \left(j_1,k_1 \right), \left(j_2,k_2 \right), \psi \in \left[0,1 \right]:\left|o^i_{j_1,k_1}-\tilde{o}^i_{j_1,k_1} \right| \leq \psi,\notag\\
&\left|o^i_{j_2,k_2}-\tilde{o}^i_{j_2,k_2} \right| \leq \psi,\ o^i_{j,k}= \tilde{o}^i_{j,k}\ \forall (j,k) \neq \left(j_1,k_1 \right), \left(j_2,k_2 \right).
\end{align}

For the occupancy data, we bound the allowed deviation on the reported occupancy, when we add or remove one vehicle. This implies we offer no privacy protection for vehicles that change the measured cumulative occupancy or the average speed excessively (the occupancy contribution of one car is proportional to the inverse of its velocity). Therefore, the occupancy time due to a single vehicle is $o_{\textrm{car}}T=\frac{l_{\textrm{car}}}{v_{\textrm{car}}}$ where $l_{car}$ is the average car length.% If changing the trajectory of a single vehicle changes the measured occupancy by $\psi$, its velocity must be $v_{\textrm{car}}=\frac{l_\textrm{car}}{\psi T}$. This velocity, even for a $7$-meter-long car is $\frac{0.84}{\psi} km/h$ that means the vehicle occupies the sensor line for $\psi$.

Considering the occupancy $0\leq o^i_{j,k}\leq 1$, the adjacency relation~\eqref{eqn:occadjac} results in a high sensitivity and the corresponding standard Gaussian perturbation mechanism leads to unreliable occupancy pseudo-measurements, especially when the number of single-loop detectors in the road increases. Instead of using the occupancy measurements to estimate the density directly, we reconsider the mode pseudo-measurement model presented in Algorithm. \ref{alg:1} from a differential privacy perspective, that is, we examine how model \eqref{eqn:model1} behaves when the trajectory of a single vehicle changes. Thus, let us rewrite \eqref{eqn:model1} as
\begin{equation} \label{eqn:model2}
M^i_k=\begin{cases}
F \qquad \text{if}\ \mathbf{1}_{T_F-T_C}\left((\Phi^i_k,y^i_k )\right)=1 \\ 
C \qquad \text{if}\ \mathbf{1}_{T_C-T_F}\left((\Phi^i_k,y^i_k )\right)=1 \\
M^i_{k-r}\ \text{if}\ \left[ \prod^{r-1}_{s=0} \mathbf{1}_{T_C \cap T_F}\left((\Phi^i_{k-s},y^i_{k-s}) \right) \right]\times\\
\qquad\qquad\qquad \mathbf{1}_{\bar{T}_C \cup \bar{T}_F}\left((\Phi^i_{k-r},y^i_{k-r}) \right)=1, \ r>0
\end{cases} 
\end{equation}
by replacing the flow measurements $\phi^i_k$ with the flow pseudo-measurements $\Phi^i_k$. By changing the trajectory of a single vehicle, we have
\begin{equation}
\label{eqn:adjmodel}
\tilde{M}^i_k=\begin{cases}
F \qquad \text{if}\ \mathbf{1}_{T_F-T_C}\left((\tilde{\Phi}^i_k,\tilde{y}^i_k) \right)=1\\ 
C \qquad \text{if}\ \mathbf{1}_{T_C-T_F}\left((\tilde{\Phi}^i_k,\tilde{y}^i_k )\right)=1\\
\tilde{M}^i_{k-r}\ \text{if}\ \left[ \prod^{r-1}_{s=0} \mathbf{1}_{T_C \cap T_F}\left((\tilde{\Phi}^i_{k-s},\tilde{y}^i_{k-s}) \right) \right]\times\\
\qquad\qquad\qquad\mathbf{1}_{\bar{T}_C \cup \bar{T}_F}\left((\tilde{\Phi}^i_{k-r},\tilde{y}^i_{k-r} \right)=1, \ r>0.
\end{cases} 
\end{equation}
Defining $\tilde{y}^i_{k}-y^i_{k}=\Delta y^i_{k} $, $\tilde{\Phi}^i_{k}-\Phi^i_{k}=\Delta \Phi^i_k$, and according to the adjacency relations defined in \eqref{eqn:adjac} and \eqref{eqn:occadjac}, we have
\begin{align} \label{eqn:sensbounds}
&\forall i \in \left[1,S \right], \exists k_1, k_2: \left| \Delta y^{i}_{k_1} \right| \leq \dfrac{\psi}{g \lambda^i}, \left| \Delta y^{i}_{k_2} \right| \leq \dfrac{\psi}{g \lambda^i},\Delta y^i_{k}=0,\notag\\
&\left| \Delta \Phi^i_{k_1} \right| \leq \dfrac{1}{T \lambda^i} , \left| \Delta \Phi^i_{k_2} \right| \leq \dfrac{1}{T \lambda^i}, \Delta \Phi^i_{k}=0 \ \forall i \neq i_0.
\end{align}
Consequently, we characterize all possible cases of mode switching due to changing the trajectory of a single vehicle in the following lemma.
\begin{lem} \label{lemprivacy}
For the sets $T_F$ and $T_C$ defined in \eqref{T_F} and \eqref{T_C}, and all flows $\Phi^i_k$, we have
\begin{align} \label{eqn:datayeghabl1}
\begin{array}{lcr}
\quad \mathbf{1}_{T_F}\left((\Phi^i_k,y^i_k)\right) \mathbf{1}_{T_C}\left((\tilde{\Phi}^i_k,\tilde{y}^i_k)\right)=0\\
\&\ \mathbf{1}_{T_C}\left((\Phi^i_k,y^i_k)\right) \mathbf{1}_{T_F}\left((\tilde{\Phi}^i_k,\tilde{y}^i_k)\right)=0,\ \textrm{if}\ \Phi^i_k \notin \left[ \alpha,q_{\max}\right]
\end{array}
\end{align}
where 
\begin{align*}
\alpha=\min&\left\{\frac{e^{-\zeta(g)}\left(\rho_{\max}-\frac{1}{T \lambda^i w}\right)-\frac{\psi}{g\lambda^i}}{\frac{e^{\zeta(g)}}{v_f}+\frac{1}{e^{\zeta(g) }w}},\right.\\
&\quad \left.\frac{e^{-\zeta(g)}\rho_{\max}-\frac{ e^{\zeta(g)}}{T \lambda^i v_f}-\frac{\psi}{g \lambda^i}}{\frac{e^{\zeta(g)}}{v_f}+\frac{1}{e^{\zeta(g) }w}}\right\}.
\end{align*}
\end{lem}
This lemma shows that, for the flow pseudo-measurement $\Phi^i_k \leq \alpha$, changing the trajectory of a single vehicle does not affect the outcome of mode measurement model \eqref{eqn:model2}. Accordingly, we now develop our privacy-preserved mode measurement model. Based on Lemma \ref{lemprivacy}, we first define the sets 
\begin{align*}
PT_F=\left\{\left(\Phi^i_k,y^i_k\right):\left|\log\left[\tfrac{\Phi^i_{k}}{v_f}\right]-\log\left[y^i_k\right]\right|\leq\zeta(g),\right.\\
\left.\Phi^i_k \in [0,\alpha)\ \forall i,k\right\}    
\end{align*} 
\begin{align*}
PT_C=\left\{\left(\Phi^i_k,y^i_k\right):\left|\log\left[\rho_{\max}-\tfrac{\Phi^i_{k}}{w}\right]-\log\left[y^i_k\right]\right|\leq\zeta(g),\right.\\
\left.\Phi^i_k \in [0,\alpha)\ \forall i,k\right\}
\end{align*}
corresponding to the pseudo-flow $\Phi^i_k$ satisfying our private truncation in free or congested modes. Defining $\bar{PT}_C$ and $\bar{PT}_F$ as the complement sets of $PT_C$ and $PT_F$, respectively, we obtain the following privacy-preserved mode measurement model:
\begin{equation}
\label{eqn:model3}
M^i_k=\begin{cases}
F\qquad \text{if}\ \mathbf{1}_{PT_F-PT_C}\left((\Phi^i_k,y^i_k )\right)=1\\
C \qquad \text{if}\ \mathbf{1}_{PT_C-PT_F}\left((\Phi^i_k,y^i_k )\right)=1\\
M^i_{k-r}\ \text{if}\ \left[\prod^{r-1}_{s=0} \mathbf{1}_{PT_C \cap PT_F}\left((\Phi^i_{k-s},y^i_{k-s}) \right) \right]\times\\
\qquad \qquad \quad \mathbf{1}_{\bar{PT}_C \cup \bar{PT}_F}\left((\Phi^i_{k-r},y^i_{k-r}) \right)=1,\ r>0
\end{cases} 
\end{equation}

\begin{figure}
\centering
\includegraphics[width=0.95\linewidth]{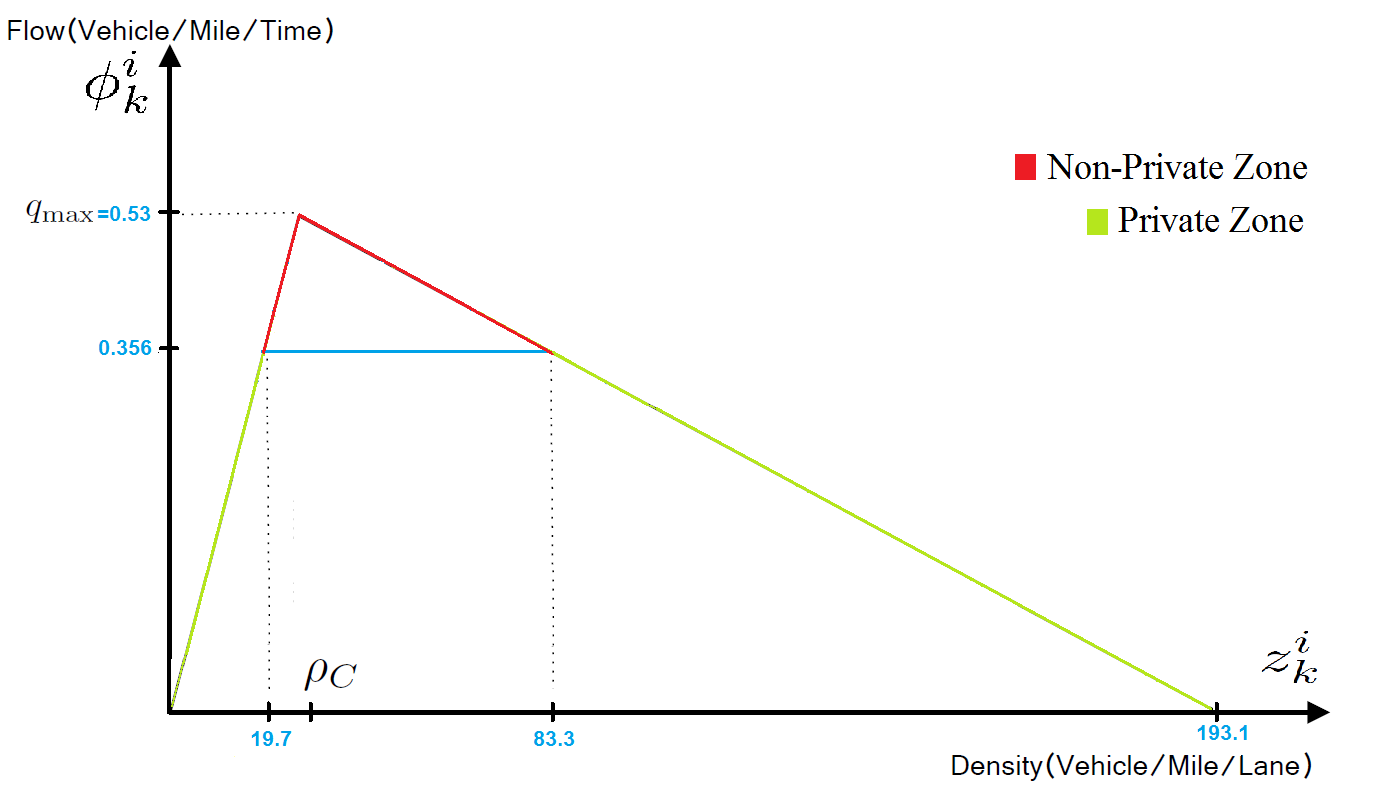}
\caption{The traffic mode in the Private zone is robust to any change in the trajectory of a single vehicle.}
\label{fig:privatelwr}
\end{figure}

As shown in Fig. \ref{fig:privatelwr}, we can divide the triangular fundamental diagram into two zones, called the \enquote{Private} zone and the \enquote{Non-Private} zone. For illustrative purposes, we depict the lines corresponding to these zones with different colors in Fig. \ref{fig:privatelwr}. The Private zone represents the flow intervals for which the traffic mode can be estimated uniquely, and also changing the trajectory of a single vehicle does not affect the mode estimation. The diagram is depicted for a four-lane road and the parameters related to the fundamental diagram and the model are $v_f=65$ mph, $w=11.6$ mph, $\rho_{\max}=193$ vehicles/mile/lane, $g=20$ feet, and $ \zeta(g) =0.51$. We also set $\psi=0.25$ that is sufficiently large to protect the privacy of individuals, i.e, all the vehicles that cross the sensor line faster than $3$ km/h, assuming the vehicles are at least $7$ meters long.
Our model for flows in the Non-Private zone estimates the traffic mode at each sensor location based on the previous estimated mode for which the flow is in the Private zone. Adopting this strategy minimizes the possibility of privacy leakage while it still provides a meaningful observation signal to specify the mode. We summarize our proposed procedure for obtaining privacy-preserved density pseudo-measurements in Algorithm \ref{alg:2}.

\begin{algorithm}
    \caption{Privacy-preserved density measurement.} \label{alg:2}
\begin{algorithmic}[1]
 \State Perturb the flow measurements \eqref{eqn:flows} to obtain the differentially-private flow pseudo-measurements $\Phi^i_k=\phi^i_k+n^i_k$.
 \State Based on historical data, choose a base g-factor, e.g, 20 feet, and an upper-bound error $\zeta(g)$.
 \State Set the maximum deviation $\psi$ in two adjacent occupancy data. Note: making $\psi$ too large spoils the mode estimation with the goal of protecting the privacy of excessively slow vehicles.
 \State Specify the corresponding Private and Non-Private zones based on Lemma \ref{lemprivacy}.
 \State Calculate $m^i_k=\frac{F}{C}$ based on the mode measurement model \eqref{eqn:model3}.
 \State Filter $m^i_k$ through the HMM filter to obtain the actual mode $s^i_k$ used to invert the fundamental diagram.
 \State Calculate
 \begin{equation*}
z^i_k=z^{i+1}_{k}=\begin{cases}
\frac{\Phi^i_k}{v_f}\qquad\qquad\ \text{if}\quad s^i_k=F\\
\rho_{\max}-\frac{\Phi^i_k}{w}\quad \text{if}\quad s^i_k=C.\\
\end{cases}
\end{equation*}
\end{algorithmic}
\end{algorithm}

\section{Traffic State Estimation} \label{section6} 

\begin{figure}
\centering
\includegraphics[width=0.65\linewidth]{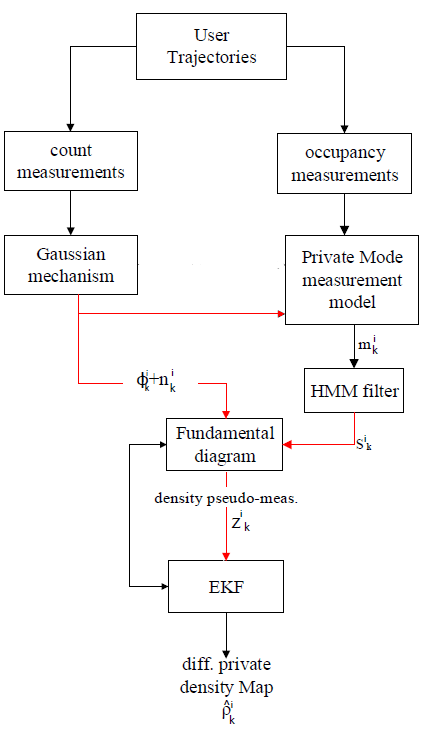}
\caption{Architecture of our differentially-private traffic estimator. The red arrows represent differentially-private signals, i.e., perturbed flow pseudo-measurements from vehicle counts. The private mode estimation is done using both counts and occupancy measurements.}
\label{fig:architect}
\end{figure}

In this section, we present the overall architecture of our differentially-private traffic state estimator, and illustrate its performance on the Mobile Century dataset \cite{Herrera2010568}. Fig. \ref{fig:architect} illustrates the overall architecture of our privacy-preserving traffic estimator. The EKF assimilates the dynamic traffic model \eqref{eqn:discrete} and the density pseudo-measurements $z^i_k$ obtained from the occupancy and count measurements.   

The differential privacy guarantee provided by this architecture is the sum of guarantees provided by the Gaussian mechanism and our privacy-preserving mode measurement model. Recalling Theorem \ref{thm: Gaussian mech}, we specify the privacy guarantee provided by our mode measurement model in the following lemma and subsequent theorem.
\begin{lem} \label{vii1}
Consider $D=\left\{c^i_{1,k},..,c^i_{\lambda^i,k},o^i_{1,k},..,o^i_{\lambda^i,k}\right.$ $\left.\forall i,k\right\}$ collecting the count and occupancy data with adjacency relations defined by \eqref{eqn:adjac} and \ref{eqn:occadjac}. Let $d$, $d'$ be two adjacent elements in $D$ with $K$ rows and define $E=\{E_i:E_i=e_1\times e_2\times ...\times e_K,e_i\in[0,\alpha)\textrm{ or }[\alpha,q_{\max}]\ \forall i=1,...,2^K\}$ with $\alpha$ defined in Lemma \ref{lemprivacy} specifying the Private zone. Then, for our privacy-preserving mode measurement mechanism $M$ and the flow pseudo-measurement $\Phi$, we have 
\begin{equation} \label{esbatlem}
M(d)=M(d')\ \textrm{if}\ \Phi(d),\Phi(d')\in E_i,\forall d, d' \in D, \forall i=1,...2^K.
\end{equation}
\end{lem}

\begin{thm} \label{vii2}
The privacy-preserving mode estimation mechanism defined in \eqref{eqn:model3} is ($\epsilon$, $\delta$)-differentially private. 
\end{thm}

Finally, in light of Theorem \ref{jamemech}, the DP guarantee for the overall architecture is $(2\epsilon, 2\delta)$, the aggregation of $(\epsilon, \delta)$-DP for the mode measurement and $(\epsilon, \delta)$-DP for the flow pseudo-measurement.

\section{Results and Discussion}

\begin{figure}
\centering
\includegraphics[width=0.7\linewidth]{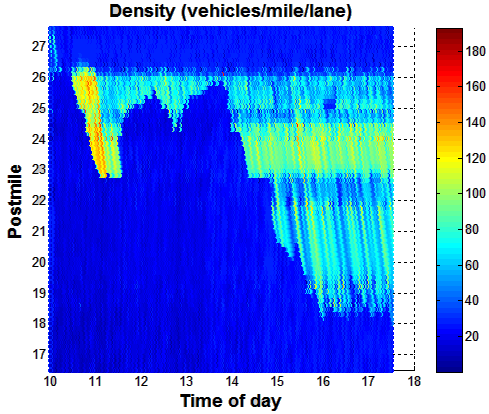}
\caption{Real-time density map reconstruction with $(\log(2), 0.05)$-DP guarantee presented based on our approach.}
\label{privateme}
\end{figure}

\begin{figure}
\centering
\includegraphics[width=0.7\linewidth]{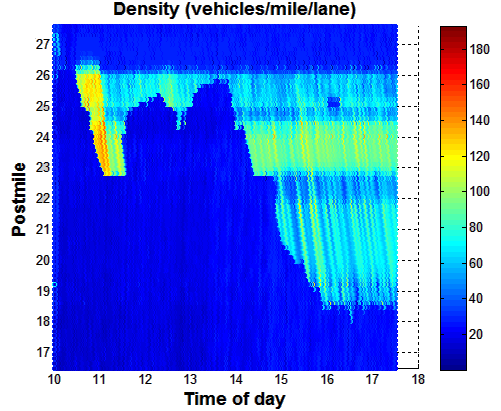}
\caption{Real-time density map reconstruction with $(\log(4), 0.1)$-DP guarantee presented based on our approach.}
\label{privateme2}
\end{figure}

Figs. \ref{privateme} and \ref{privateme2} show examples of $(\log(2), 0.05)$ and  $(\log(4), 0.1)$-differentially-private maps, respectively, based on our designed privacy-preserving traffic estimator. The complete map is built using 10 out of the 27 sensors placed at different locations on four lanes of the US I-880 highway.

\begin{figure}
\centering
\includegraphics[width=0.7\linewidth]{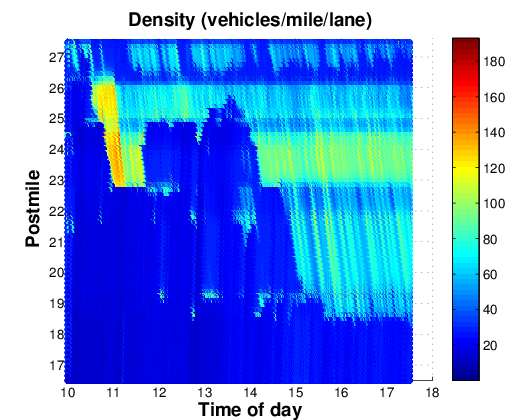}
\caption{Real-time density map reconstruction with $(10+\log(2), 0.05)$-DP guarantee presented in \cite{asli}.}
	\label{privjerom}
\end{figure}

Our proposed approach improves over the state-of-the-art in terms of privacy guarantee and has immediate applications in providing privacy preservation for traffic monitoring in long highways. It also improves the reliability of the reconstructed density maps by significantly reducing the instances of incorrect switching between the modes. Comparison of the three maps in Figs. \ref{privateme}, \ref{privateme2}, and \ref{fig:nonprivatehmmfilterasli-eps-converted-to2} shows that we can obtain strong $(\log(4), 0.1)$ or even stronger $(\log(2), 0.05)$ privacy guarantees using the proposed approach with negligible degradation in estimation performance. However, our approach may estimate the mode of the flows in the Non-Private zone with error, mainly due to the delay introduced by our mode measurement model. For example, assume that the traffic flow increases and the road becomes congested. The mode of the traffic based on our privacy-preserving mode measurement model will be free until the flow reenters the Private zone. In this case,  according to Fig. \ref{fig:privatelwr}, we may encounter up to $60$ (vehicles/mile/lane) errors in our density maps. The upper-bound of these errors can be tightened by decreasing the value of the parameter $\psi$. For example, $\psi=0.1$ can decrease the upper-bound of this errors to $40$ units, but it also weakens the privacy guarantee. One effective way to improve the mode measurement when the flow is in the Sensitive zone is to use this model:
\begin{equation}
\forall\ \Phi^i_k \in [\alpha, q_{\max}]:
\begin{cases}
F\ \ \textrm{if}\ \ \Phi^i_k-\Phi^i_{k-1}>0\\ 
C\ \ \textrm{if}\ \ \Phi^i_k-\Phi^i_{k-1}<0.
\end{cases} 
\end{equation}     
This model takes advantage of the fact that the traffic flow dynamic is either decreasing or increasing given the congested or free modes, respectively. However, at this point, it unfortunately appears unwieldy for use with a more advanced mechanism.

\section{Conclusion} \label{sec:conclusion}

We presented new methods for monitoring traffic while protecting the privacy of individual users whose data is used to estimate the traffic state at a particular location of a road. To this end, we used a macroscopic hydrodynamic model to analyze the dynamics of the variables involved. Our approach was different from earlier privacy-preserving methods used for location-based services as we focused on releasing aggregated data, such as traffic speed and density, while ensuring that the privacy of individuals is not compromised.

\section*{Appendix I\\Background on Differential Privacy} \label{sec:back4}
\setcounter{section}{1}

The basic problem setting in differential privacy for a statistical dataset is as follows. Suppose a curator is in charge of a statistical dataset, which consists of records of many individuals, and an analyst sends a query request to the curator to obtain some aggregate information about the whole dataset. Without any privacy concerns, the curator can simply apply the query function to the dataset, compute the query output, and send the result to the analyst. However, to protect the privacy of individual data in the dataset, the curator should use a randomized query-answering mechanism such that the probability distribution of the query output does not differ significantly whether or not any individual record is in the
dataset.

\subsection{Definition of Differential Privacy} \label{section: DP def}

Formally, we start by defining a symmetric binary relation, denoted by $\Adj$, on the space of datasets of interest $\D$, which is used to define
what it means for two datasets to differ by the data of a single individual. For any $d,d'$ subsets of $\D$ , we have $\Adj(d,d')$ if and only if we can obtain the signal $d'$ from $d$ simply by adding or subtracting the data of one user. Mechanisms that are differentially private necessarily randomize their outputs, in such a way that they satisfy the following property.
\begin{defn} \label{def: differential privacy original}
Let $\D$ be a space equipped with a symmetric binary relation denoted by $\Adj$ and let $(\R, \mathcal M)$ be a measurable space. Let $\epsilon, \delta \geq 0$. A mechanism $M: \D \times \Omega \to \R$ is $(\epsilon, \delta)$-differentially private for $\Adj$ if for all $d,d' \in \D$ such that $\Adj(d,d')$, we have
\begin{align} \label{eq:dp}
\Prob(M(d) \in S) \leq e^{\epsilon} \Prob(M(d') \in S) + \delta, \;\; \forall S \in \mathcal{M}. 
\end{align}
If $\delta=0$, the mechanism is said to be $\epsilon$-differentially private. 
\end{defn}
This definition quantifies the allowed deviation for the output distribution of a differentially private mechanism, when a single individual is added or removed from a dataset. If the inequality fails, a leakage, known as $(\epsilon,\delta)$ breach, takes place. This means that the difference between the prior and posterior distributions is tangible.
 
%In some cases, we index the mechanism by the query $q$ of interest, writing $M_q$.
The choice of the parameters $\epsilon, \delta$ is set by the privacy policy. Typically, $\epsilon$ is taken to be a small constant, e.g., $\epsilon \approx 0.5$. The parameter $\delta$ has to be kept small as it controls the probability of certain significant losses of privacy, e.g., when a zero probability event for $d'$ becomes an event with positive probability for $d$ in (\ref{eq:dp}).

One fundamental property of the notion of differential privacy that is used in this work is the characterization of differential privacy under adaptive composition. The following theorem shows that the privacy degrades under composition to the sum of the differential privacy parameters of each access.
\begin{thm} \label{jamemech}
Consider $ M_1,...M_r$ as $r$ mechanisms on the space $\D$ where $M_i$ is ($\epsilon_i$,$\delta_i$)-differentially private. The mechanism $M = (M_1,....,M_r)$, which outputs $\left(M_1(d),.....M_r(d)\right)$ for $d \in D$, is $\left(\sum^r_{i=1} \epsilon_i,\sum^r_{i=1} \delta_i\right)$-differentially private.
\end{thm}

\subsection{A Basic Differentially Private Mechanism} \label{section: basic mech}

A trivial mechanism that discards all the information in a dataset is obviously private but not useful. In general, one has to trade off privacy for utility when answering specific queries. Below, we recall a basic mechanism that can be used to answer queries in a differentially private way.
We are only concerned with queries that return numerical answers, i.e., here a query is a map $q: \D \to \R$ where the output space $\R$ equals $\mathbb R$ is equipped with a norm denoted by $\| \cdot \|_\R$ and the $\sigma$-algebra $\mathcal M$ on $\R$ is taken to be the standard 
Borel $\sigma$-algebra. %, denoted $\mathcal R^k$. 
The following quantity plays an important role in the design of differentially private mechanisms~\cite{10.1007/11681878_14}.

\begin{defn} \label{defn: sensitivity}
Let $\D$ be a space equipped with an adjacency relation $\Adj$. The sensitivity of a query $q: \D \to \R$ is defined as $\Delta_\R q := \max_{d,d':\Adj(d,d')} \|q(d) - q(d')\|_\R$. In particular, for $\R = \mathbb R$ equipped with the $1$-norm $\|x\|_1 = |x|$, we denote the $\ell_1$ sensitivity by $\Delta q= \max_{d,d':\Adj(d,d')} |q(d) - q(d')|$.
\end{defn}
%\begin{enumerate}
    %\item 
    
We now present two generic mechanisms that guarantee $\epsilon$- and ($\epsilon,\delta$)-differential privacy.

1) \emph{Laplace Mechanism:} This mechanism modifies an answer to a numerical query by adding zero-mean noise distributed according to a Laplace distribution. Recall that the Laplace distribution with mean zero and scale parameter $b$, denoted by $\textrm{Lap}(b)$, has density $p(x;b)=\frac{1}{2b}\exp\left(-\frac{|x|}{b}\right)$ and variance $2b^2$. Moreover, for $\omega \sim \textrm{Lap}(b)$, we have $E[|\omega|]=b$, and $\Prob(|\omega|\geq tb)=e^{-t}$. 
\begin{thm}	\label{thm: Lap mech}
Let $q: \D \to \mathbb R$ be a query, $\epsilon>0$. The mechanism $M_q: \D \times \Omega \to \mathbb R$ defined by $M_q(d) = q(d) + w$ with $w \sim \textrm{Lap}(b)$, where $b \geq \frac{\Delta q}{\epsilon}$, is $\epsilon$-differentially private.
\end{thm}

2) \emph{Gaussian Mechanism:} A differentially-private mechanism proposed in \cite{Dwork06_DPgaussian} modifies an answer to a 
numerical query by adding iid zero-mean Gaussian noise.
Recall the definition of the $\mathcal Q$-function $\mathcal Q(x) := \frac{1}{\sqrt{2 \pi}} \int_x^{\infty} e^{-\frac{u^2}{2}} du$. Hence, we have the following theorem~\cite{Dwork06_DPgaussian, LeNyDP2012_journalVersion}.
%\vspace{0.1cm}
\begin{thm}	\label{thm: Gaussian mech}
Let $q: \D \to \mathbb R$ be a query and $\epsilon>0$. The Laplace mechanism $M_q: \D \times \Omega \to \mathbb R$ defined by $M_q(d) = q(d) + w$ with $w \sim \mathcal N\left(0,\sigma^2 \right)$, where $\sigma \geq \frac{\Delta q}{2 \epsilon}(K + \sqrt{K^2+2\epsilon})$ and $K = \mathcal Q^{-1}(\delta)$, is $(\epsilon,\delta)$-differentially private.
\end{thm}

We define $\kappa_{\delta,\epsilon} = \frac{1}{2 \epsilon} (K+\sqrt{K^2+2\epsilon})$. Therefore, the standard deviation $\sigma$ in Theorem \ref{thm: Gaussian mech} can be written as $\sigma(\delta,\epsilon) = \kappa_{\delta,\epsilon} \Delta q$. It can be shown that $\kappa_{\delta,\epsilon}$ behaves roughly as $\mathcal{O}(\ln\frac{1}{\delta})^{1/2}/\epsilon$. For example, to guarantee $(\epsilon,\delta)$-differential privacy with $\epsilon = \ln(2)$ and $\delta = 0.05$, the standard deviation of the introduced Gaussian noise has be about $2.65$ times the $\ell_1$-sensitivity of $q$.
%\end{enumerate}

\subsection{Utility Measure: Usefulness}

We propose to construct a novel dataset access mechanism whose results can be released to the public and be useful, that is, its output well approximates the true query results. We formally define the notion of utility below~\cite{Blum:2008:LTA:1374376.1374464}.
\begin{defn}(Usefulness Definition). \label{defn:useful}
A dataset access mechanism $M_q$ is ($\gamma, \zeta$)-useful, if with probability $1 - \zeta$, for every dataset $d \subseteq \D$, we have $|M_q(d)- q(d)| \leq \gamma$.
\end{defn}
\begin{thm}
The Laplace Mechanism~\ref{thm: Lap mech} is $(\frac{\Delta q}{\epsilon} \ln \frac{1}{\zeta},\zeta)$-useful \cite{Chan2011}. Equivalently, the Laplace Mechanism~\ref{thm: Lap mech} is $(\gamma,\frac{1}{exp(\frac{\gamma \cdot \epsilon}{\Delta q})})$-useful.
\end{thm} 
\begin{thm}
The Gaussian Mechanism~\ref{thm: Gaussian mech} is $(\gamma,\frac{2\cdot \gamma}{\sigma(\delta,\epsilon) \cdot \Delta q})$-useful.
\end{thm}

\section*{Appendix II\\Extended Kalman Filter}

Consider the following non-linear stochastic state-space system:
\begin{align}
\label{eqn:ONLS}
x_{k+1}&=F(x_k)+\omega_k, \quad k \in \mathbb{Z}_{+}  \\
y_k&=H(x_k)+\nu_k
\end{align}
where $ x_0 \sim \mathcal{N} (0 ,\Sigma)$ is independent of the system disturbance process $\omega$ and the observation noise process $\nu$. We also assume 
\begin{align*}
\left[ \begin{array}{lcr} \omega \\ \nu \end{array} \right] \sim \ \mathcal{N} \left( \left[ \begin{array}{lcr}0 \\ 0\end{array} \right],\left[\begin{array}{lcr}Q \ \ 0 \\ 0 \ \ R \\\end{array} \right] \right).
\end{align*}
A popular approach to the stochastic state estimation for system \eqref{eqn:ONLS} is the extended Kalman filter~\cite{simon06}. Subject to the assumption that $F$ and $H$ have continuous first-order partial derivatives, one may recursively employ the Taylor series expansion of $F$ and $H$ to obtain linear approximations of the system dynamics and observations processes in the neighborhood of the time-varying trajectory $x_k, k \in \mathbb{Z}_{+} $.
Henceforth, we adopt this assumption without any further comment. Using a first-order approximation of $F(x_k)$, the estimated state $\hat{x}_k$ can be obtained via the following conditioning and prediction steps.\\
Conditioning step:
\begin{align*}
\hat{x}_k= x_{k|k-1}+V_k H^T_k\left[H_kV_kH^T_k+R\right]^{-1}\left(y_k-H(x_{k|k-1})\right)
\end{align*}
Prediction step:
\begin{align*}
V_{k+1}= F_k V_k F^T_k-F^T_kH^T_k \left[H_kV_kH^T_k+R\right]^{-1} H_kF_k+Q
\end{align*}
where
\begin{align*}
&x_{k+1|k}=F(\hat{x}_k), \quad V_0= \Sigma,\\
&F_k=\left[\frac{\partial F(x)}{\partial x} \right]_{x=\hat{x}_k},\quad H_k=\left[\frac{\partial H(x)}{\partial x} \right]_{x=x_{k|k-1}}.
\end{align*}

\section*{Appendix III\\Proofs}
\setcounter{subsection}{0}

\subsection{Proof of Lemma \ref{thm:nonprivatezone}}

%\begin{proof}
We know that
\begin{equation}
\label{eqn:modelbug1}
 e ^{-\zeta(g)} \frac{\phi^i_{k}}{v_f}\leq y^i_k \leq e ^{\zeta(g)} \frac{\phi^i_{k}}{v_f}, \ \ \forall (\phi^i_k,y^i_k ) \in T_F \end{equation}
\begin{align} \label{eqn:modelbug2}
e^{-\zeta(g)}\left(\rho_{\max}-\frac{\phi^i_{k}}{w}\right) \leq y^i_k \leq e ^{\zeta(g)} \left(\rho_{\max}-{\frac{\phi^i_{k}}{w}}\right),\notag\\
\forall (\phi^i_k,y^i_k ) \in T_C.
\end{align}
Hence, we have
\begin{align*}
T_F \cap T_C= \left\{(\phi^i_k,y^i_k):\ e^{-\zeta(g)} \frac{\phi^i_{k}}{v_f} \leq e^{\zeta(g)} \left(\rho_{\max}-\frac{\phi^i_{k}}{w}\right)\right.\\
\left.\& \ e^{-\zeta(g)}\left(\rho_{\max}-\frac{\phi^i_{k}}{w}\right) \leq e^{\zeta(g)} \frac{\phi^i_{k}}{v_f}  \right\}
\end{align*}
and, by solving the inequalities for $\phi^i_k$, we get
\begin{align*}
\mathbf{1}_{T_F}&\left((\phi^i_k,y^i_k)\right) \mathbf{1}_{T_C}\left((\phi^i_k,y^i_k)\right)=1\\
&\textrm{iff}\ \phi^i_k \in \left[\dfrac{w v_f \rho_{\max}}{w e^{2 \zeta(g)} +v_f},\dfrac{w e^{2 \zeta(g)} v_f \rho_{\max}}{w+e^{2 \zeta(g)} v_f} \right].
\end{align*}
%\end{proof}

\subsection{Proof of Lemma \ref{lemprivacy}}

%\begin{proof}
Considering Lemma \ref{thm:nonprivatezone}, \eqref{eqn:modelbug1}, and \eqref{eqn:modelbug2}, we have
\begin{align*}
&\mathbf{1}_{T_F}\left((\Phi^i_k,y^i_k)\right) \mathbf{1}_{T_C}\left((\tilde{\Phi}^i_k,\tilde{y}^i_k)\right)=1\ \textrm{iff}\notag\\
&e^{-\zeta(g)}\dfrac{\Phi^i_{k}}{v_f} \leq \left[e^{\zeta(g)}\left(\rho_{\max}-{\dfrac{\Phi^i_{k}+\Delta \Phi^i_k}{w}}\right)\right]-\Delta y^i_{k}\ \&\notag\\
&\left[e^{-\zeta(g)}\left(\rho_{\max}-\dfrac{\Phi^i_{k}+\Delta \Phi^i_k}{w}\right)\right]-\Delta y^i_{k} \leq e^{\zeta(g)} \dfrac{\Phi^i_{k}}{v_f}
\end{align*}
and
\begin{align} \label{eqn:switching}
&\mathbf{1}_{T_C}\left((\Phi^i_k,y^i_k)\right) \mathbf{1}_{T_F}\left((\tilde{\Phi}^i_k,\tilde{y}^i_k)\right)=1\ \textrm{iff}\notag\\
&e^{-\zeta(g)}\left(\rho_{\max}-\dfrac{\Phi^i_{k}}{w}\right) \leq \left[e ^{\zeta(g)} \dfrac{\Phi^i_{k}+\Delta \Phi^i_k}{v_f}\right]-\Delta y^i_{k}\ \&\notag\\
&\left[e ^{-\zeta(g)}\dfrac{\Phi^i_{k}+\Delta \Phi^i_k}{v_f}\right]-\Delta y^i_{k} \leq e ^{\zeta(g)}\left(\rho_{\max}-{\dfrac{\Phi^i_{k}}{w}}\right)
\end{align}
or equivalently
\begin{align*}
&F \rightarrow C\ \ \textrm{if}\ \ \Phi^i_k \in A\\ 
&C \rightarrow F\ \ \textrm{if}\ \ \Phi^i_k \in B\\
&A=\left[\dfrac{\left[e^{-\zeta(g)}\left(\rho_{\max}-\dfrac{1}{T \lambda^i w}\right) \right]-\dfrac{\psi}{g \lambda^i}}{\dfrac{e^{\zeta(g)}}{v_f}+\dfrac{1}{e^{\zeta(g) }w}},q_{\max}\right]\\
&B=\left[\dfrac{e^{-\zeta(g)}\rho_{\max}-\dfrac{ e^{\zeta(g)}}{T \lambda^i v_f}-\dfrac{\psi}{g \lambda^i}}{\dfrac{e^{\zeta(g)}}{v_f}+\dfrac{1}{e^{\zeta(g) }w}},q_{\max}\right]
\end{align*}
where we limit the maximum of each interval by $q_{\max}$ to prevent any privacy leakage. The minimizations are also over the corresponding parameters based on \eqref{eqn:sensbounds}. Finally, we obtain the proof considering that
\begin{align} \label{eqn:datayeghabl}
\begin{array}{lcr}
\quad\mathbf{1}_{T_F}\left((\Phi^i_k,y^i_k)\right) \mathbf{1}_{T_C}\left((\tilde{\Phi}^i_k,\tilde{y}^i_k)\right)=0\\
\&\ \mathbf{1}_{T_C}\left((\Phi^i_k,y^i_k)\right) \mathbf{1}_{T_F}\left((\tilde{\Phi}^i_k,\tilde{y}^i_k)\right)=0,\ \textrm{if}\ \Phi^i_k \notin A \cup B.
\end{array}
\end{align}
%\end{proof}

\subsection{Proof of Lemma \ref{vii1}}

%\begin{proof}
Given a pair of adjacent data elements, say, $d$ and $d'$, if their pseudo-flows, $\Phi(d)$ and $\Phi(d')$, are in the same zone, the mode measurement model will result in identical outputs, since the model always estimates the mode with respect to the flows in the Private zone. Hence, the model automatically ignores any change in occupancy measurements due to adding or removing a single vehicle.
%\end{proof}

\subsection{Proof of Theorem \ref{vii2}}

%\begin{proof}
Defining $\mathbb{\chi} = \{F,C \}^K$, $\forall d,d'\in D$ and $s \in \chi$, we have
\begin{align*}
\mathbb{P}&(M(d)\in s)\\
&=\sum^{2^k}_{i=1} \left[ \mathbb{P} \left(M(d)\in s \ | \ \Phi(d)\in E_i \right) \mathbb{P} \left(\Phi(d)\in E_i \right)\right]\\
&=\sum^{2^k}_{i=1} \left[ \mathbb{P} \left(M(d')\in s \ | \ \Phi(d')\in E_i \right) \mathbb{P} \left(\Phi(d)\in E_i \right)\right]
\end{align*}
where the latter equality is in light of \eqref{esbatlem}. The flow $\Phi(d)=\phi(d)+n$ is the output of a Gaussian mechanism and is ($\epsilon$, $\delta$)-differentially private. Therefore, we have
\begin{multline*}
\mathbb{P}\left(\Phi(d)\in E_i\right)=\frac{1}{(2\pi\sigma^2)^{k/2}}\times\\
\int_{E_i}e^{-\frac{\left\|u-\phi(d')\right\|^2}{2\sigma^2}}e^{\frac{2(u-\phi(d'))^T (\phi(d)-\phi(d'))-\left\|\phi(d)-\phi(d')\right\|^2}{2 \sigma^2}}du\\ 
\leq e^{\epsilon}\mathbb{P}\left(\Phi(d')\in E_i \right)+\frac{1}{(2\pi\sigma^2)^{k/2}}\int_{E_i}e^{-\frac{\left\|u-\phi(d)\right\|^2}{2\sigma^2}}\times\\
\mathbf{1}\left\{2\left(u-\phi(d')\right)^T\left(\phi(d)-\phi(d')\right) \geq \left\|\phi(d)-\phi(d')\right\|^2\right.\\
\left.+ 2\epsilon \sigma^2\right\}du.
\end{multline*}
The last integral term defines a measure that is bounded by $\delta$ (for more details, see the proof of Theorem 3 in~\cite{LeNyDP2012_journalVersion}). Let $A$ be the flow area specified by the indicator function. Then, we have
\begin{align*}
&\frac{1}{(2 \pi \sigma^2)^{k/2}}\int_{E_i}e^{\frac{\left\|u-\phi(d)\right\|^2}{2 \sigma^2}}\times\\
&\mathbf{1}\left\{2(u-\phi(d'))^T (\phi(d)-\phi(d')) \geq \left\|\phi(d)-\phi(d')\right\|^2\right.\\
&\hspace{6.5cm}\left.+ 2\epsilon\sigma^2\right\}du\\
&=\mathbb{P}(\Phi(d)\in \left[ A \cap E_i \right])\\
&=\mathbb{P}(\Phi(d)\in A)\ \mathbb{P}(\Phi(d)\in E_i\ |\ \Phi(d)\in A).
\end{align*}
As we know $\sigma^2 = \left\|\phi(d)-\phi(d') \right\|^2 \kappa^2_{\epsilon,\delta}$, it is easy to show that 
\begin{align*}
\mathbb{P}\left(\Phi(d)\in A\right)&=\delta\\
\mathbb{P}(\Phi(d)\in \left[A \cap E_i \right])&=\delta\, \mathbb{P}(\Phi(d)\in E_i \ | \ \Phi(d)\in  A) .
\end{align*}
Therefore, we have
\begin{align*}
&\mathbb{P}(M(d)\in s) \leq \sum^{2^k}_{i=1} \mathbb{P} \left(M(d')\in s \ | \ \Phi(d')\in E_i \right)\times\\
&\qquad\qquad \left[ e^{\epsilon} \mathbb{P} \left(\Phi(d')\in E_i \right) +\delta \ \mathbb{P}(\Phi(d)\in E_i \ | \ \Phi(d)\in  A)  \right]\\
&=e^{\epsilon} \mathbb{P} \left(M(d')\in s \right)\\
&+\delta \sum^{2^k}_{i=1} \mathbb{P} \left(M(d')\in s \ | \ \Phi(d')\in E_i \right)\mathbb{P}(\Phi(d)\in E_i \ | \ \Phi(d)\in  A)\\
&=e^{\epsilon} \mathbb{P} \left(M(d')\in s \right)\\
&+\delta \sum^{2^k}_{i=1}  \mathbb{P} \left(M(d)\in s \ | \ \Phi(d)\in E_i \right)\mathbb{P}(\Phi(d)\in E_i \ | \ \Phi(d)\in A).
\end{align*}
The last sum is bounded by $1$ as it is over mutually-exclusive events $\Phi(d)\in E_i$ conditioned on a single event $\Phi(d)\in A$. This concludes the proof.
%\end{proof}

\section*{Acknowledgment}

We thank Dr Jerome Le Ny for providing invaluable guidance and support throughout this work.

\end{document}